\documentclass[12pt]{article}
\usepackage{epsf}

\def\ltap{\raisebox{-.4ex}{\rlap{$\sim$}} \raisebox{.4ex}{$<$}}

\def\ltap{\raisebox{-.4ex}{\rlap{$\sim$}} \raisebox{.4ex}{$<$}}

\begin{document}
\begin{titlepage}
\today          \hfill
\begin{center}
\hfill    LBNL-42812 \\
\hfill    UCB-PTH-99/03\\
{\large \bf
Extra Dimensions and the Muon Anomalous Magnetic Moment}
\footnote{This work was supported in part
by the Director, Office of Energy
Research, Office of High Energy and Nuclear Physics,
Division of High
Energy Physics of the U.S. Department of Energy
under Contract
DE-AC03-76SF00098.
The author was also supported by NSERC.}
\vskip .25in

M. L. Graesser
\footnote{email: graesser@thwk8.lbl.gov}

{\em Theoretical Physics Group, 
    Lawrence Berkeley National Laboratory\\
    and \\
    Department of Physics, 
      University of California\\
    Berkeley, California 94720}
\end{center}

\begin{abstract}
It has been proposed recently that the 
scale of strong gravity can be very close to the weak scale. 
Dimensions of sizes anywhere from $\sim $mm to $\sim $TeV$^{-1}$ 
can be populated by bulk gravitons, vector bosons and fermions. 
In this paper the one-loop correction of these bulk particles 
to the muon magnetic moment (MMM)
are investigated. 
In all the scenarios 
considered here it is found that the natural value for the 
MMM is $O(10^{-8}-10^{-9})$.
One main result is that 
the contribution of 
each Kaluza$-$Klein graviton to the MMM is remarkably finite. 
The bulk graviton loop implies a limit of $\sim 400$ GeV on the scale 
of strong gravity. 
This could be pushed up to $\sim 1-2$ TeV, even in the case of six extra 
dimensions, if the BNL E821 experiment reaches 
an expected sensitivity of $\sim 10^{-9}$. 
Limits on 
a bulk $B-L$ gauge boson are interesting, but still allow for 
forces $10^6-10^7$ times stronger than gravity at mm$^{-1}$ distances.
The correction of a bulk right-handed 
 neutrino to the MMM in one recent proposal for 
generating small Dirac neutrino masses is considered in the 
context of a two Higgs doublet model, and is found to 
be close to $10^{-9}$. The contributions 
of all these bulk particles to the MMM are (roughly) independent of 
both the total number of extra dimensions and the dimension of the 
subspace occupied by the bulk states. Finally, 
limits on 
 the size of ``small''                      
compact dimensions 
gotten from the MMM and atomic 
parity violation are determined and compared.
\end{abstract}

\end{titlepage}

\renewcommand{\thepage}{\arabic{page}}
\setcounter{page}{1}

\section{Introduction.}

In the past year there has been a 
remarkable proposal by Arkani-Hamed, Dimopoulos and Dvali (ADD)
for solving the hierarchy problem \cite{nima0,nima00,nima1}. 
In their vision the scale of strong 
quantum gravity is lowered to the weak scale. A  
weak-scale valued Planck scale $M_*$ and Newton's constant are 
reconciled by allowing  
gravity to propagate in $n$ extra {\it large} 
compact dimensions. An application 
of Gauss' Law quantifies the relationship between 
$\bar{M}^2_{PL}=(8\pi G_N)^{-1}$, $M_*$, and the volume of the compact 
space $V_n=R^n$, to be  
\begin{equation}
2\bar{M}^2_{PL}=M^{n+2}_* R^n.
\end{equation}
In the case 
of 2 extra dimensions and the phenomenologically 
preferred value $M_* \sim$TeV, 
$R \sim $mm and future short-distance experiments 
may observe a transition in Newton's Law from a $r^{-2}$ to 
$r^{-4}$ force law. In the case of 6 extra dimensions and 
$M_* \sim $TeV, $R^{-1} \sim$MeV, 
which is   
large compared to $M_* \sim $TeV. The non-observation 
of Kaluza-Klein (KK) towers for the Standard Model (SM) particles, 
for example, 
requires that the SM particles not propagate in the large 
extra dimensions. This requires that they be fixed to a 
3$-$dimensional wall, with a thickness smaller than 
$\sim O(\hbox{(700 GeV)}^{-1})$. 
This may be accomplished using purely field-theoretic techniques \cite{nima0}, 
or by using the D-branes of string theory \cite{nima00}.  
This remarkable proposal allows for many new interesting phenomena to 
be discovered at future experiments.
These include  
dramatic model-independent signals at high-energy colliders 
from bulk graviton production \cite{wells,peskin}, signatures of the 
spin-2 structure of 
the KK gravitons \cite{hewett}, distortions in the Drell-Yan energy 
distribution \cite{hewett}, and enhancements in the $f^+ f^-$, 
$WW$, $ZZ$, 
$\gamma \gamma$ production 
cross-sections at linear colliders \cite{hewett,agashe}, as 
well as other interesting collider signatures \cite{sridhar} and 
implications for gravitational processes at high energies \cite{shrock}.  
Future colliders may also 
discover  
Kaluza-Klein and  
string excitations of the Standard Model particles 
\cite{antoniadis1,antoniadis2,nima00}. 
The existence of either {\it large} or TeV$^{-1}$ sized 
dimensions also opens new territory
for model-building. In an important early 
paper \cite{antoniadis1} by Antoniadis, 
the author advocates the existence of  TeV$^{-1}$ sized dimensions  
to break supersymmetry at low-energies. 
Other prospects include new ideas for obtaining gauge coupling 
unification \cite{dienes,kakushadze}, suppressing proton decay, and 
flavor physics\cite{nima3,nima4}.  
Existing 
phenomena already strongly 
constrain $M_*$. These include astrophysical processes (such as SN1987A in 
the case $n=2$) \cite{nima1}, 
bulk graviton production \cite{wells,peskin}, and 
$WW, ZZ$ production at LEP2 \cite{agashe}. 

In this paper the bulk contributions at the one-loop level to 
the anomalous magnetic moment of the muon (MMM) are computed. 
These include contributions from bulk gravitons, bulk $B-L$ gauge bosons, 
and, in a two Higgs doublet model, those from bulk 
 right-handed (RH) neutrinos. 
The correction to the MMM is found to be $\sim m^2_{\mu}/M^2_*$, and 
is therefore in a first approximation {\it independent of n}, 
or, 
in the case of bulk fermions or vector bosons, 
independent of 
 the 
{\it dimension of the subspace} inhabited by the bulk states. 

Now 
it may seem strange to compute a one-loop correction to the MMM 
when a correction from a TeV-scale 
suppressed operator may be present at tree-level. The important 
point is that 
the tree-level operator contains an unknown coefficient, whereas 
the one-loop 
corrections are present, calculable and  
essentially model-independent. They thus represent a lower value to 
the full bulk correction to the MMM.

In this paper it is found that the ``expected'' one-loop 
correction to 
the MMM is typically $O(10^{-8}-10^{-9})$ for values of 
$M_*$ from $\sim$ 400 GeV to $\sim O(\hbox{TeV})$. 
For comparison, this is the same magnitude as 
the correction expected in 
low-energy supersymmetry at large $\tan \beta$ \cite{moroi}.
The bulk correction to the MMM 
should be compared to its extremely well-measured 
value of \cite{pdg}      
\begin{equation}
a^{exp}_{\mu}=\frac{g-2}{2} =(116592.30 \pm 0.8) )\times10^{-8}.
\label{error}
\end{equation}
The BNL experiment (E821) hopes to lower the
error to $40 \times 10^{-11}$ \cite{BNL}, a factor of $\sim 20$ 
below the existing sensitivity.
The prediction in the SM is \cite{marciano,kinoshita}
\begin{equation}
a^{SM}_{\mu}=116591.739(154)\times 10^{-8},
\end{equation}
and is in very good agreement with the current measurement.
The theoretical error is dominated by the hadronic contributions. 
The error in the lowest order hadronic contribution is $\sim 60$ ppm of 
$a_{\mu}$, but there is hope that this error can be reduced to 
$\sim 0.5$ ppm \cite{BNL,takeo}. Another large source of theoretical 
error is the hadronic 
``light-on-light'' 
scattering correction to $a_{\mu}$. This error   
is estimated to be $\sim 0.15$ ppm \cite{kinoshita}. 
These translate into an 
expected theoretical error of $\sim 5 \times 10^{-10}$.

A main result of this paper is the computation of the one-loop 
correction from bulk gravitons and radions  
to the MMM. The 
corrections from a single KK graviton and KK radion are
found to each be miraculously finite. The full one-loop bulk contribution 
is gotten by summing over all 
the KK states, and results in a correction to the MMM 
which is essentially 
model-independent and independent of $n$, 
depending only on the scale of strong gravity. 
Requiring that the correction is smaller than the present  
experimental error implies,  
in the case $n=6$, for instance, that the lower limit on $M_*$  
is competitive with limits gotten 
from other physical processes.
It is also found that as a general rule
the $a_{\mu}$ constraint provides a comparable  
limit to $M_*$ for {\it larger} numbers of extra dimensions.
This is in contrast to astrophysical
or terrestrial constraints gotten from the direct production of gravitons, 
where the effect of a physical process
with characteristic energy scale $E$ 
typically decouples as $\sim (E/M_*)^n$. 
As such, the $a_{\mu}$ constraint provides
complementary information. 

Bulk RH neutrinos were proposed in Ref. \cite{nima4} to generate
small Dirac neutrino masses, and can naturally account for the required
mass splitting needed to explain the atmospheric neutrino anomaly.
It is found that the correction to the MMM from the tower of 
RH neutrinos in a two Higgs doublet model 
is close to the future experimental sensitivity.

It is also found that while the constraint on the  
$B-L$ gauge coupling is quite impressive, 
 a $B-L$ vector boson can still mediate isotope-dependent forces
$10^{6}-10^{7}$ times stronger than gravity at sub-mm distance scales.

Finally, KK excitations of the SM gauge bosons are also probed 
by the MMM and atomic parity violation measurements. It is found 
that the stronger limits on the size of extra ``small'' dimensions
are gotten from the 
atomic parity violation experiments. 
The limit is $\sim 1$ TeV $(\sim 3.6$ TeV $)$ for KK excitations into 
1 (2) extra dimensions. 

The outline of the paper is as follows. Section 1 discusses some 
machinery necessary to perform the one-loop graviton correction to the MMM. 
Section 2 discusses the results and limits gotten from the one-loop 
graviton computation. Section 3 discusses the bulk $B-L$ and KK photon 
contributions to the MMM, and concludes with a discussion of the KK $Z_0$ 
correction to atomic parity violation measurements. 
Section 4 discusses the correction 
to the MMM 
in a recent proposal for generating small Dirac masses within the 
context of a two Higgs-doublet model. 
 
\section{Preliminaries and Some Machinery.}
\label{section1}

The effective field theory approach 
described by Sundrum in Ref. \cite{sundrum}
allows one to determine the couplings of the higher dimensional 
gravitons to some matter stuck on a wall. 
The interactions are constrained 
by local 4-dimensional Poincare 
invariance (which interestingly enough, 
arise from the coordinate 
reparameterisation invariance of the wall embedding) 
and the full $(4+n)-$dimensional 
general coordinate invariance, 
as well as any local gauge symmetries on 
the wall. At lowest order in $1/M_*$, 
all the higher dimensional 
gravitons interact only through the 
induced metric on the wall. Since the interaction 
of the induced metric with on-the-wall matter is fixed by $4-$D general 
coordinate invariance, the 
lowest order couplings of the KK gravitons 
to the on-the-wall matter are fixed by the ordinary graviton coupling
to matter, 
and are obviously universal. This is 
important, since the interactions of matter to the KK gravitons 
are therefore determined by replacing
$h_{\mu \nu} \rightarrow \sum_n h^{(n)}_{\mu \nu}$ in
the ordinary gravity-matter interactions.
With $g=\eta+\kappa h$, where 
$\kappa =\sqrt{32 \pi G_N}$, 
$h_{\mu \nu}(x)=\sum_n h^{(n)}_{\mu \nu}(x)$, 
the gravitons couple to the conserved symmetric 
stress-energy tensor with normalisation
\begin{equation}
-\kappa h^{\mu \nu} T_{\mu \nu} /2.
\label{tuv}
\end{equation}

There is one subtlety though. The $4-$D decomposition of 
the $4+n$ dimensional graviton results in a KK tower of  
spin-2 gravitons, $n-1$ vector bosons, and $n(n-1)/2$ spin-0 
bosons. The vector bosons correspond to fluctuations in $g_{i \mu}$. 
In the linearised theory they couple only to $T^{i \mu}$, 
which vanishes if the wall has no momentum 
in the extra dimensions. The zero mode of the vector bosons acquires a mass 
$\sim M^2_* /M_{PL}$ from the spontaneous breaking of translation invariance 
in the extra dimensions \cite{nima5}. The zero mode of one of the 
spin-0 bosons corresponds to 
fluctuations in the size of the compact dimensions (the ``radion'') and 
couples to $T^{\mu}_{\mu}$. It acquires 
a mass $\sim 10^{-3}$eV to MeV from the dynamics responsible for 
stabilising the size of the extra dimensions \cite{nima5}. 
The remaining spin-0 fields do not couple to 
$T^{\mu} _{\nu}$ \cite{wells,rules}.

The end result is that only the tower of KK gravitons and radions 
couple to the matter on the wall. 
Since $T^{\mu}_{\mu}$ 
 is proportional to the fermion masses, it is important to 
include the radion and its tower of KK states in computing the 
correction to $g-2$. 
A careful computation \cite{wells,rules} determines the normalisation 
of the coupling of the radion to $T^{\mu}_{\mu}$. 
The correct prescription is then to 
replace 
\begin{equation}
h_{\mu \nu}(x)\rightarrow\sum_n \left(h^{(n)}_{\mu \nu}(x)
-\sqrt{\omega} \eta_{\mu \nu} \phi^{(n)}(x) \right) .
\label{expansion}
\end{equation}
For a canonical normalisation of $\phi$, $\omega=(n-1)/3(n+2)$.
The couplings of the KK tower of 
gravitons and radions to matter is then gotten 
from the ordinary graviton coupling to matter by using the 
substitution in Eqn. \ref{expansion}.

Next I present some of the machinery which is needed to perform the 
one-loop bulk graviton correction to the MMM. The reader 
is referred to References \cite{veltman} for a far more careful 
treatment of this subject, and for early quantum gravity computations. 
The recent References \cite{wells,rules} present a clear 
discussion of the proper procedure for determining the bulk graviton 
couplings to matter.
 
The Lagrangian coupling 
gravity to on-the-wall photons and fermions is
\begin{equation}
\sqrt{-g} {\cal L}=\sqrt{-g} 
\left(-\frac{1}{4}g^{\mu \rho} g^{\nu \sigma}
F_{\mu \nu} F_{\rho \sigma}
-\frac{1}{2 \xi}(g^{\mu \nu} D_{\mu} A_{\nu})^2+ 
\bar{\psi} i \gamma^a e^{\mu}_a D_{\mu}\psi
-m \bar{\psi} \psi \right),  
\label{L1}
\end{equation}
with   
$D_{\mu} A_{\nu}
=\partial _{\mu}  A_{\nu}-\Gamma^{\rho} _{\mu \nu} A_{\rho}$, 
$\Gamma$ is the Christoffel symbol, 
and $\xi$ is the $U(1)$ gauge fixing parameter.
The $e^{\mu}_a$ fields are the vielbeins satisfying
$g^{\mu \nu}=e^{\mu}_a \eta^{a b} e^{\nu} _b$ and
$\eta^{ab}=e^{\mu a} g_{\mu \nu} e^{\nu b}$.
The fermionic covariant 
derivative is $D_{\mu}=\partial _{\mu}-i e A_{\mu}
-\frac{i}{2} \omega^{ab}_{\mu} \sigma_{ab}$, with
$\sigma_{ab}=\frac{i}{4} [\gamma_a,\gamma_b]$. Finally,
the spin connection
is
\begin{equation}
\omega^{ab} _{\mu}
=\frac{1}{2}\left(e^{\nu a}(\partial _{\mu} e^{b} _{\nu}
-(\mu \leftrightarrow \nu))-(a \leftrightarrow b) \right)
-\frac{1}{2} 
e^{\nu a}e^{\rho b} e_{\mu d} \left(\partial _{\nu} e^{d} _{\rho}
-(\nu \leftrightarrow \rho)\right).
\end{equation}
Inserting $g_{\mu \nu}=\eta_{\mu \nu}+ \kappa h_{\mu \nu}$,
$e_{\mu a}=\eta_{\mu a}+\kappa h_{\mu a}/2$, 
and expanding Eqn. \ref{L1} to $O(h)$ gives
\begin{eqnarray}
{\cal L} &=& {\cal L}(h=0)+ 
\frac{1}{2 } \kappa h^{\mu \nu} 
\left(\eta^{\rho \sigma} F_{\mu \rho} F_{\nu \sigma}
-\frac{1}{4} \eta_{\mu \nu} F^{\alpha \beta} F_{\alpha \beta} \right)
\nonumber \\ 
& & +\frac{1}{2 \xi} \kappa (\partial \cdot A)\left( 
-\frac{1}{2} h (\partial \cdot A) 
+  h^{\mu \nu} (\partial _{\mu} A_{\nu}+\mu \leftrightarrow \nu) 
+ A^{\alpha} \eta^{\mu \nu}
\left(h_{\alpha \mu, \nu}+h_{\alpha \nu,\mu}-h_{\mu \nu, \alpha} 
\right) \right)  \nonumber \\  
& & + \frac{\kappa}{2}
\left( (h \eta^{\mu \nu}- h^{\mu \nu})\bar{\psi} i
\gamma_{\mu} D_{\nu} \psi-m h \bar{\psi} \psi
+\frac{1}{2}(\partial _{\mu} h-\partial ^{\nu} h_{\mu \nu})
\bar{\psi} i \gamma^{\mu} \psi ) \right).
\label{Lh1}
\end{eqnarray}
The minimal coupling of gravity to the matter stress-energy tensor, 
Eqn. \ref{tuv}, 
is gotten by integrating Eqn. \ref{Lh1} by parts to remove the 
derivatives on $h$.

The propagator for a massive spin-2 particle, with mass $m_N$, is 
\begin{equation}
D(k^2,m^2_N)_{\mu \nu, \rho \sigma} = 
\frac{1}{2} \frac{i P_{\mu \nu, \rho \sigma}}{k^2-m^2_N}.
\label{massiveprop1}
\end{equation}
The tensor structure in the numerator is  
\begin{eqnarray}
P_{\mu \nu, \rho \sigma} &=&\left(\eta_{\mu \rho}-
\frac{k_{\mu} k_{\rho}}{m^2_N}\right) \left(\eta_{\nu \sigma}-
\frac{k_{\nu} k_{ \sigma}}{m^2_N}\right)+\left(\eta_{\mu \sigma}-
\frac{k_{\mu} k_{\sigma}}{m^2_N}\right)\left(\eta_{\nu \rho}-
\frac{k_{\nu} k_{\rho}}{m^2_N}\right) \\ \nonumber 
& &-\frac{2}{D-1} 
\left(\eta_{\mu \nu} -\frac{k_{\mu} k_{\nu}}{m^2_N} \right)
\left(\eta_{\rho \sigma}-\frac{k_{\rho} k_{\sigma}}{m^2_N}\right) 
\label{massiveprop2}
\end{eqnarray} 
and is uniquely fixed by several requirements \cite{veltman}. 
The first is that the 
graviton is the excitation of the metric, so $P$ is symmetric 
in $\mu \leftrightarrow \nu$ 
and $(\mu \nu) \leftrightarrow (\rho \sigma)$. 
Next, unitarity implies that 
on-shell $P$ is positive-definite and that 
the spin-1 and spin-0 components 
are projected out. Finally, the unitarity requirement 
$P(k^2=m^2_N,m^2_N)_{\mu \nu, \rho \sigma}=
\sum _S \epsilon ^{S} _{\mu \nu} \epsilon ^{S} _{\rho \sigma}$, where 
$\epsilon^{S}_{\mu \nu}$ is one of 5 polarisation tensors, 
fixes the normalisation of $P$.

It will be useful to compare the ordinary massless 
graviton one-loop correction to $a_{\mu}$ \cite{massless} to 
the massive spin-2 
one-loop correction 
in the limit $m_N \rightarrow 0$. 
The massless spin-2 propagator is 
required for the former computation. In the harmonic gauge it is 
\begin{equation}
D(k^2)_{\mu \nu, \rho \sigma} =
\frac{1}{2} \frac{i P_{\mu \nu, \rho \sigma}}{k^2}, \hbox{ }
P(k^2)_{\mu \nu, \rho \sigma}= \eta_{\mu \rho} \eta_{\nu \sigma} 
+ \eta_{\mu \sigma} \eta_{\nu \rho} 
-\frac{2}{D-2} \eta_{\mu \nu} \eta_{\rho \sigma} .
\end{equation}
It will be important to note that the coefficient of 
$\eta_{\mu \nu} \eta_{\rho \sigma}$ in the numerator 
of the graviton propagator differs between the massless and massive cases. 

Each Kaluza-Klein (KK) mode $n$ of a bulk graviton 
contributes to $a_{\mu}$. 
It is conveinent to express the contribution of a Feynman diagram $i$, 
with internal KK mode $n$, to
$a_{\mu}$ as 
\begin{equation}
\Delta a^{(n,i)} _{\mu} = \Delta ^{(n)} (g-2)_i/2 
=G_N m^2_F {\cal A}^{(n)}_i /  2 \pi.
\label{amplitude}
\end{equation}
The fermion 
mass is $m_F$.  
The total
correction to $a_{\mu}$ is then 
gotten by summing over all the KK modes 
and Feynman diagrams.
The final result is 
\begin{equation}
 \frac{G_N}{2 \pi}  m^2_F \sum_n {\cal A}^{(n)} _F(m^2_F,m^2_N), 
\hbox{ } \hbox{ }
\hbox{ } 
{\cal A}^{(n)}_F={\cal A}^{(n)}_1+\cdots + {\cal A}^{(n)}_K, 
\hbox{for }K \hbox{ diagrams}.  
\label{tR}
\end{equation} 
This formula can be trivially modified to include bulk fermion 
or vector boson corrections to the MMM.
Since the coupling of matter to 
the KK graviton, vector boson or fermion modes is universal 
\footnote{I assume for simplicity that the SM fermions do not 
have KK excitations.}, the 
only dependence of $a^{(n)}_i$ on the KK quantum number $n$ enters 
through the mass of the KK mode, $m_N=n/R$. 
Replacing the 
sum in Eqn. \ref{tR} with an integral gives the final result 
\begin{equation}
\Delta a_{\mu}=\frac{G_N R^n}{2 \pi}  
m^2_F \frac{\pi^{n/2}}{\Gamma[n/2]}
\int dm^2 (m^2)^{(n/2-1)} {\cal A}_F(m^2_F,m^2) \hbox{ }.
\label{fR}
\end{equation}
This is a good approximation since the mass splitting is tiny: 
$\sim 10^{-3} \hbox{ eV}$ for $n=2$, $\sim 5 \hbox{ MeV}$ for $n=6$.

In evaluating the one-loop contributions to the MMM, 
the dominant 
correction  
will occur from those states in the loop with masses close to the 
cutoff. This is a direct consequence of the large multiplicity of 
states at large KK masses. The total correction to the MMM is then 
well-approximated by substituting into Eqn. \ref{fR} the large $m_N$ limit 
of ${\cal A}_F$, and cutting off the (power-divergent) integral 
over $m^2_N$ at KK masses $m_N=M_*$. 
For the cases considered in this paper, 
in the large $m_N$ limit ${\cal A}_F \rightarrow c/(m^2_N)^W$ 
with $W=0$ or $1$. Then 
using the relation $(4 \pi G_N)^{-1}=M^{n+2}_* R^n$, 
a good approximation to Eqn. \ref{fR} is, 
for $W=0$ say, 
\begin{equation}
\Delta a_{\mu}=\frac{\lambda}{8 \pi^2}c 
\frac{2 \pi^{n/2}}{n \Gamma[n/2]} 
\frac{m^2_F}{M^2_*}.
\label{cutoffsum}
\end{equation}
The unknown coefficient $\lambda$ is a gross parameterisation 
of our ignorance about what is 
regulating the integral, and can only be computed in a finite 
theory of quantum gravity, e.g. string theory. The expectation is 
that $\lambda \sim O(1)$ since, for example, 
in string theory 
the amplitude ${\cal A}_F$ becomes exponentially 
suppressed at KK masses larger than the string scale.
So whereas $\lambda$ is an unknown $O(1)$ coefficient, 
the limits on $M_*$ gotten from the bulk graviton 
loop scale as $\lambda^{1/2}$ and are therefore not so 
sensitive to its value. Finally, it is consistent 
to keep 
the factors of $\pi^{n/2}$, etc. in  Eqn. \ref{cutoffsum}, 
since this factor counts the degeneracy of 
states,  which can only increase in the full theory.

\section{Bulk Gravity.}

{\it a. \underline{Spin-2 Graviton}}

\begin{figure}
\centerline{\epsfxsize=0.6\textwidth \epsfbox{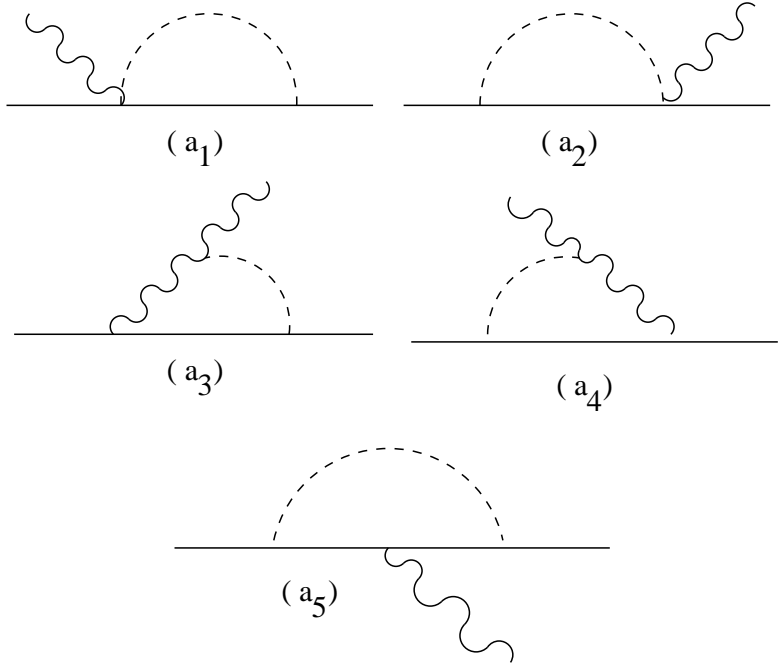}}
\caption{Bulk gravity contribution at the one-loop level
to the anomalous magnetic
moment. Dashed lines denote either
bulk spin-2 graviton or
bulk spin-0 radion. Wavy and solid lines denote on-the-wall
photons and fermions,
respectively.}
\protect\label{extradfig1}
\end{figure}

Using the Lagrangian in Eqn. \ref{Lh1} the Feynman 
rules for the graviton and fermion, photon interactions can be 
gotten\footnote{See references \cite{wells,rules} for an  
explicit presentation of the Feynman rules.}. 
Using this, and 
the massive graviton propagator, 
Eqn. \ref{massiveprop1}, it is straightforward 
although tedious to compute the one-loop correction to $a_{\mu}$. 
The Feynman diagrams relevant to the calculation are shown in Fig. 
\ref{extradfig1}. Those 
terms in the numerator of the graviton propagator 
containing two or more $k_{\alpha}$s do not contribute to $a_{\mu}$ 
at this order in $G_N$ 
because of the gravitational Ward identity. I have
also verified this by an explicit computation. 
With this in mind, naive 
power counting implies that the 
contribution of each Feynman diagram 
and for a given KK mode $n$  
to $a_{\mu}$ is logarithmically divergent in the ultraviolet. 
The explicit calculation presented below reveals though 
that these divergences 
cancel in the sum, so that the total contribution of each 
KK mode to $a_{\mu}$ 
is remarkably finite! A similar miraculous cancellation was 
observed in the ordinary (massless) graviton one-loop correction 
to $a_{\mu}$ \cite{massless}. 

In the notation of Eqn. \ref{amplitude} $({\cal A}^{(n)}_i=a^{(n)}_i)$, the 
partial amplitudes $a^{(n)}_i$ are $(D=4-\epsilon)$  
\begin{eqnarray}
\label{r1}
a^{(n)} _1=a^{(n)} _2 &=&-\frac{11}{3} \frac{1}{\epsilon}+\frac{1}{3}
+\frac{11}{6} \ln m^2_F 
  - \frac{1}{2} \int dx 
\frac{R(x,m^2_F,m^2_N)}{L(x,m^2_F,m^2_N)} A(x), \\ 
\label{r2}
a^{(n)} _3=a^{(n)} _4 &=&  \frac{4}{\epsilon} +4 - 2 \ln m^2_F - 
\int dx \left(R(x,m^2_F,m^2_N) B(x) 
 +8 \frac{m^2 _F}{L(x,m^2_F,m^2_N)} C(x) \right) , \\ 
\label{r3}
a^{(n)} _5 &=& -\frac{2}{3} 
\frac{1}{\epsilon}+\frac{5}{8}+\frac{1}{3} \ln m^2_F
-\int dx \left(R(x,m^2_F,m^2_N) D(x)
+\frac{m^2_F}{L(x,m^2_F,m^2_N)} E(x) \right)
\end{eqnarray}
where $L(x,m^2_F,m^2_N)=x^2 m^2_F +(1-x) m^2_N$ contains the only 
dependence of $a^{(n)}_i$ on the 
KK quantum number. The other functions are  
$R(x,m^2_F,m^2_N)=(2x m^2_F-m^2_N)/L(x,m^2_F,m^2_N)$, 
$A(x)=4 x-x^3 /3$, $B(x)=14 x^2/3-20 x /3$, 
$C(x)=x^2 -x^3/2$, $D(x)=3 x^2 /2-5 x^3 /3+ x^4 /2$,
and $E(x)=16 x^2 /3-20 x^3 /3 +3 x^4-x^5 /2$. 
As promised above, the sum over all the 5 diagrams is 
finite: the coefficient of 
$1/ \epsilon$ is $-22/3+8-2/3=0$! The final result 
is 
\begin{equation}
a^{(n)}_F=a^{(n)} _1+ \cdots +a^{(n)} _5
=\frac{223}{24}- \int dx \frac{2xm^2_F- m^2_N}{L(x,m^2_F,m^2_N)} H(x)
-\int dx \frac{m^2_F}{L(x,m^2_F,m^2_N)} P(x) ,
\label{rF}
\end{equation}
where $H(x)=x (1-x)(-28/3 +3 x/2 -x^2/2)$ and 
$P(x)=-x^5/2+3 x^4-44 x^3/3+64 x^2/3$. 
An interesting limit is $m^2_N \gg m^2_F$. In this 
case the last term in Eqn. \ref{rF} can be neglected, and
$a^{(n)}_F=223/24-14/3+1/2-1/8=5$. 
Note that the heavy KK states                                     
exhibit ${\it non-decoupling}$. A short comment about this 
will be made 
later.

The one-loop correction from the ordinary massless graviton 
was computed years ago in Ref. \cite{massless}.
A comparison of the massive spin-2 correction, 
Eqns. \ref{r1}, \ref{r2}, \ref{r3}, 
in the limit $m^2_N \rightarrow 0$ 
to their results 
provides a non-trivial check on the calculation.  
Naively
the contribution of each KK bulk graviton to $a_{\mu}$ 
in this 
limit should
equal the
ordinary massless graviton.
This is because the coupling of matter
to all the graviton KK modes is universal, including the zero mode.
This expectation is
incorrect though, since the tensor structure
of the massive and massless graviton
propagators are different. Essentially there are more
helicity states flowing around the massive graviton loop 
than there are in the massless graviton loop. 
Specifically, the coefficient of the
$\eta_{ \mu \nu} \eta_{\rho \sigma}$ term
in the numerator of the graviton 
propagator is $-1/(D-1)$ and $-1/(D-2)$,
respectively, and they are not the same. Also note 
that the $\eta_{ \mu \nu} \eta_{\rho \sigma}$ contribution is
proportional to the radion correction, discussed below. 
The massless
graviton correction to $a_{\mu}$
is then gotten by adding and
subtracting this ``$\eta-\eta$'' contribution 
evaluated at $m^2 _N =0$, weighted by either
$-1/(D-1)$ or $-1/(D-2)$. 
That is, the ordinary graviton correction to 
$(g-2)/2$ should be   
\begin{eqnarray}
b_i(\hbox{massless graviton}) & = &
a^{(n)} _i(m^2_N \rightarrow 0)+
\frac{1}{D-1} \tilde{a}^{(n)} _i
- \frac{1}{D-2} \tilde{a}^{(n)} _i \\ \nonumber 
 &= & a^{(n)} _i(m^2_N =0) 
-\frac{1}{D-1} \frac{1}{D-2} \tilde{a}^{(n)} _i,
\label{rel}
\end{eqnarray}
where $\tilde{a}^{n}_i$ is the contribution to $(g-2)/2$, evaluated 
at $m^2_N=0$, of the 
$\eta_{ \mu \nu} \eta_{\rho \sigma}$ 
term in the massive graviton propagator. That is, $\tilde{a}^{(n)}_i 
=\omega^{-1} r^{(n)}_i(m^2_N=0)$ where $r^{(n)}_i$ is the contribution 
of a KK radion, discussed below. 

In this limit Eqns. \ref{r1}, \ref{r2} 
and \ref{r3} reduce to 
$a^{(n)} _1=a^{(n)} _2=-11/3 \epsilon -32/9 
+11/6 \ln m^2_F$,
$a^{(n)} _3=a^{(n)} _4=4 / \epsilon +20/3-2 \ln m^2_F$,
and $a^{(n)} _5=-2/3 \epsilon -26/ 9+1/3 \hbox{ }\ln m^2_F$.
Now an explicit computation gives 
$\tilde{a}^{(n)} _1= \tilde{a}^{(n)} _2=0, \hbox{ } \tilde{a}^{(n)}_3
= \tilde{a}^{(n)} _4=-2, 
\hbox{ } \tilde{a}^{(n)} _5=3$. 
The correction from each diagram is 
evidently finite! Using Eqn. \ref{rel}
and the $m^2_N=0$ limit of 
Eqns. \ref{r1}, \ref{r2} and \ref{r3} presented 
above, one 
gets $b_1=b_2=-11/3\epsilon -32/9$, $b_3=b_4=4/ \epsilon+20/3+2/6=
4/\epsilon+7$ and $b_5=-2/3\epsilon-26/9-1/2=-2/3 \epsilon-61/18$. 
These results are in agreement
with the computation given in Ref. \cite{massless}.

{\it b. \underline{Radion}}

The contribution of the tower of KK radions to $a_{\mu}$ is gotten by 
computing the Feynman diagrams in Fig. \ref{extradfig1}, using 
the Lagrangian in Eqn. \ref{Lh1} and the expansion of the induced 
metric, Eqn. \ref{expansion}. It is remarkable that, before summing 
over the KK states, the contribution of  
each Feynman diagram to $a_{\mu}$ is finite! 
In the notation of Eqn. \ref{amplitude} $({\cal A}^{(n)}_i=r^{(n)}_i)$, 
the result is 
\begin{eqnarray}
\label{rr1}
\nonumber
r^{(n)}_1=r^{(n)}_2 &=&0 , \\ 
\label{rr2}
\nonumber
\omega ^{-1} r^{(n)}_3= \omega^{-1} r^{(n)}_4 
&=&-4- 2 \int dx \left(R(x, m^2_F,m^2_N) F(x) 
+ \frac{m^2_F}{L(x,m^2_F,m^2_N)} G(x)\right) , \\ 
\label{rr3}
\omega^{-1} r^{(n)}_5 &=&-1-3 \int dx \left(R(x, m^2_F,m^2_N) K(x)
+\frac{2}{3}\frac{m^2_F}{L(x,m^2_F,m^2_N)} G(x) \right),
\end{eqnarray}
where $\sqrt{\omega}$ is the coupling of the radion to the 
trace of the stress-energy momentum tensor, $\omega=(n-1)/3(n+2)$. 
The functions are: $F(x)=x(1-x)(3x/2+1)$, $G(x)=x^2(3 x^2/2-2x-2)$,  
and $K(x)=x^2(1-x)$.  
In the limit $m^2_N \rightarrow0$, 
$\omega ^{-1} r^{(n)}_i  \rightarrow \tilde{a}^{(n)}_i$.  
In the 
limit $m^2_N \gg m^2_F$, 
$\omega ^{-1} r^{(n)}_3=\omega ^{-1} r^{(n)}_4
\rightarrow -2$, and $\omega ^{-1} r^{(n)}_5 \rightarrow 0$. So 
in this limit $r^{(n)}_F=-4 \omega$. 

{\it c. \underline{Applications}}

To constrain $M_*$ it is useful to consider 
the cases $m_N \ll m_F$ and
$m_N \gg m_F$. In the former case 
$a^{n}_1+\cdots+a^{n}_5$ reduces to $10/3$, 
and $r^{n}_1+\cdots+r^{n}_5$ reduces to $-\omega$.
In the $m_N \gg m_F$ case $a^{n}_1+\cdots+a^{n}_5 \rightarrow 5$, 
and $r^{n}_1+\cdots+r^{n}_5 \rightarrow -4 \omega$.
In view of the final
sum over KK states, and the fact that the sum in either limit
is $O(1)$, the (by far)
dominant contribution to $a_{\mu}$ occurs for those KK modes
with large $n$, since the degeneracy of states is much greater.
In this case the integral in Eqn.
\ref{fR} over the KK number 
phase space is trivial and power divergent.
Applying the formula of Section \ref{section1}, Eqn. \ref{cutoffsum},
and $\omega=(n-1)/3(n+2)$,
the total correction to $a_{\mu}$ is
\begin{eqnarray}
\Delta a_{\mu} &=& \frac{ \lambda}{8 \pi^2} \left( 5 
-\frac{4}{3} \left(\frac{n-1}{n+2}\right)\right)
\frac{2 \pi^{n/2}}{n \Gamma[n/2]} \frac{m^2_F}{M^2_*} \nonumber \\
&=& \frac{7\lambda}{12 \pi} \left( \frac{m_{\mu}}{M_*}\right)^2 
\hbox{ , }n=2 \hbox{ , }\nonumber \\
&=& \frac{25 \pi \lambda}{288} 
\left( \frac{m_{\mu}}{M_*}\right)^2 \hbox{ , }n=6.
\label{mainresult}
\end{eqnarray}
The first contribution 
is from the tower of spin-2 KK states, and the second 
is the contribution from the tower of KK radion states. Note that 
this result scales as $M^{-2}_*$, and is {\it independent of n}.
The presence of the factor $2\pi^{n/2}/n \Gamma[n/2]$ from 
the phase space integration substantially 
enhances the correction. For instance, for $n\sim6$ this factor 
is $\sim \pi^3/6$ and compensates the loop suppression.
Requiring that 
$\Delta a_{\mu}$ is less than the current 2$\sigma$ limit 
of $2 \times 0.8 \times 10^{-8}$
gives the following 2$\sigma$ limits :
\begin{equation}
\hbox{(current limit)   } n=2: \hbox{ } \hbox{ } M_* 
> \hbox{340 GeV } \lambda^{1/2} \hbox{ },\hbox{ } n=4,6: \hbox{ }
\hbox{ } M_* > \hbox{410 GeV } \lambda^{1/2}.
\end{equation}
Note that since the limit on $M_* \sim \lambda^{1/2}$, it is not so strongly 
sensitive to the value of $\lambda$.
The limit for $n=2$ is considerably weaker than 
existing accelerator limits. For instance, 
the good agreement between existing LEP2 data and the predicted 
SM cross-section for  
$\gamma+\hbox{missing energy}$ final states constrains the 
bulk graviton production rates \cite{peskin,wells}, and implies 
$M_* > \hbox{1200 GeV}$ for $n=2$. The $a_{\mu}$ constraint 
for $n=6$ is competitive with the limit from bulk graviton production 
which for $n=6$ is 
$M_* > \hbox{520 GeV}$ \cite{wells, peskin}.

The future prospects are more promising though. The BNL E821 experiment
plans to decrease the experimental error in $a^{exp}_{\mu}$ 
by a factor of  
$\sim 20$, i.e., down to $\Delta a^{exp} _{\mu} = 
4 \times 10^{-10}$. This should be compared to the KK graviton 
and radion contribution to the MMM. Inserting the relevant numbers into 
Eqn. \ref{mainresult} gives for $n=2$ (upper) and $n=6$ (lower):   
\begin{equation}
\Delta a_{\mu} = \left( \begin{array}{c} 2 \\3 \end{array}\right) 
\times 10^{-9} \times \lambda \times 
\left(\frac{\hbox{TeV}}{M_*}\right)^2 .
\end{equation}
So an $O(10^{-9})$ correction to the MMM is not unreasonable. 

Finally, the final formulas for the correction to MMM, and the 
non-decoupling of the KK states, can be understood by computing the 
one-loop diagram in the full theory. 
Recall that the correction to the MMM requires a helicity flip. 
This pulls out a factor of $p_{\alpha} \gamma^a p_a$, 
lowering the degree of divergence by two. Then 
naive power counting gives 
\begin{equation}
\Delta a_{\mu} 
\sim \frac{m^2_F}{M^{n+2}_*} \int dk k^{n+3}k^{-2} k^{-1} k k^{-2} 
\sim m^2_F \frac{\Lambda^n}{M^{n+2}_*} \sim \frac{m^2_F}{M^2_*},
\end{equation}
for $\Lambda\sim M_*$. 
The other factors appearing in the integrand 
are also easily understood. The first factor $k^{n+3}$ is 
the $4+n$ dimensional phase space factor. The factors of 
$k^{-2}$ and $k^{-1}$ are from the the graviton and muon propagators. 
Finally, the third factor counts the 
momentum factors at the graviton vertices; for example, in 
Fig.1(a) the graviton--fermion--fermion interaction contains $\sim k$. 
This naive explanation then reproduces Eqn. \ref{mainresult}.
The KK states exhibit non-decoupling since their coupling to matter 
is stronger at higher energies.
 
\newpage

\section{Bulk Vector Bosons}

{\it a. \underline{\it Bulk Vector Bosons}}

\begin{figure}
\centerline{\epsfxsize=0.5\textwidth \epsfbox{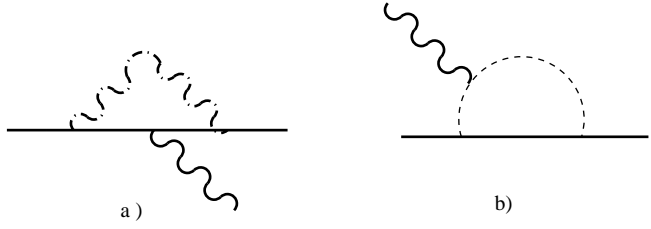}}
\caption{One-loop correction to the anomalous magnetic
moment from: a) abelian bulk vector bosons (dashed wavy line),
b) charged Higgs (dotted line) and bulk RH neutrinos.
Wavy and
solid lines denote on-the-wall photons and fermions,
respectively. }
\protect\label{extradfig2}
\end{figure}

In this Section I compute one-loop correction of a bulk $X$ gauge boson 
to the MMM. 
A $U(1)_X$ gauge field in the bulk that couples to leptons 
contributes to $a_{\mu}$ by the one-loop diagram in Fig. \ref{extradfig2}a.
It could be a gauged $B-L$, or
the higher KK modes of the photon, for example.  
As with the bulk gravitons, the bulk vector boson interactions 
are constrained by the measured value of $a_{\mu}$. In this case  
the limits are more model-dependent. Each KK mode 
couples universally to leptons with an unknown 4-dimensional 
gauge coupling $g_X$.
This bulk vector boson also does not have to live in the 
full $n$-dimensional bulk. It could instead 
live in a $p-$dimensional subspace with volume $V_p$.
It will be shown though, that those regions of parameters 
resulting in $O(10^{-8}-10^{-9})$ corrections to the MMM will 
also be probed by future short-distance force experiments.  

While the limit from $a_{\mu}$ implies that the gauge coupling 
must be microscopic (see Eqn. \ref{glimit} below), it is also quite 
natural for the $4-$dimensional coupling to be small \cite{nima1}. 
In the full $4+p$ dimensional gauge theory the gauge coupling between a  
vector boson and fermions 
is $g^{(p)}_X /M^{p/2}_*$. Confining the 
fermions to a $3-$dimensional wall, by either a topological 
defect or just using a $p-$dimensional delta function, 
 and performing the usual KK decomposition 
of the vector boson results in a universal coupling $g_X$ 
of each KK mode 
to fermions. It also gives  
\begin{equation}
\alpha_X =\alpha^{(p)}_X /(M^p_* V_p),
\label{finestructure}
\end{equation}
the relation 
between the fine structure constants in the 4-D effective theory 
and the fundamental $p-$dimensional theory.  
As a result, it is very natural for the vector boson to be very weakly 
coupled. For example, in the case $p=n$, 
$\alpha_X=\alpha^{(n)}_X M^2_* /\bar{M}^2_{PL}$, independent of $n$. In 
this case $\alpha_X \sim 2 \times 10^{-30} \times \alpha^{(n)}_X \times 
(M_* /\hbox{3 TeV})^2$.  
   
This also makes it very natural for the $X$ boson to be extremely 
light. Note 
that even if a $B-L$ gauge boson is lighter than the proton (as will 
be the case here), the proton remains stable  
since the $B-L$ gauge boson does not 
carry $B-L$ charge.
These facts were pointed out in Ref. \cite{nima1}, where it was  
argued that a $B-L$ vector boson could mediate forces at the 
sub-mm distance scale $10^7-10^8$ times stronger than gravity. 
To avoid conflicts 
with these short-distance force experiments, the mass of the zero mode 
of a new $U(1)_X$ 
coupled to baryons 
must be larger than $\sim 10^{-4} \hbox{eV}$. This requires 
that the gauge symmetry 
is spontaneously broken in the bulk by the vev $v$ of 
some scalar field $\phi$. The mass term for the gauge field in the 
full theory, written using 4-D canonically normalised fields, is 
\begin{equation}
\int d^{4+p}x \frac{1}{2} \frac{g^2_{4+p}}{M^p_*} \frac{\phi^2}{V_p} 
\frac{ {A_X}_{\mu}}{\sqrt{V_p}} \frac{ A^{\mu}_X}{\sqrt{V_p}}.
\label{gaugebosonmass}
\end{equation}
Inserting the vev, the usual KK expansion for $A_X$, integrating 
over the $p-$volume, and using Eqn. \ref{finestructure}  gives 
\begin{equation}
m_X =g_X v
\label{fmass}
\end{equation}
for the mass of the zero mode. If instead the scalar field 
$\phi$ is confined to a distant 3$-$D wall, then in Eqn. \ref{gaugebosonmass} 
I should replace $\phi^2/V_p \rightarrow \phi^2$ and insert
the approximation \cite{nima4} $\phi(x,y)\sim {\delta^{(p)}(y)} \phi(x)$.
This results in the same expression for the mass, Eqn. \ref{fmass}.  
Consistency of this field theoretic approach 
requires that the vev must satisfy $v < M_*$. This implies an 
upper bound of $g_X M_*$ to the mass of the zero mode. 
When this is combined with the constraint on $g_X$ from 
$a_{\mu}$, an interesting {\it upper bound} to the mass of the 
zero  
mode is gotten. 
The masses of the other KK states are   
\begin{equation}
m^2_N=g^2_X v^2+ \frac{n^2}{R^2}.
\label{massgb}
\end{equation}

An evaluation of Fig. \ref{extradfig2}a gives the contribution of one 
KK state to $a_{\mu}$. The result is finite and 
\begin{equation}
\Delta a^{(n)} _{\mu}=\frac{\alpha_X }{\pi} \int dx 
\frac{m^2_F}{L(x,m^2,m^2_N)} x^2 (1-x) .
\label{g2gb}
\end{equation}
The normalisation of the $U(1)_X$ charge is chosen such 
that the $X$ charge of a lepton is 1. 
The function $L(x,m^2,m^2_N)$ is 
defined in the previous section. In the limit $m_N \rightarrow 0$ 
Eqn. \ref{g2gb} reduces to the famous result 
from Schwinger \cite{schwinger}. 

I now want to consider performing the sum over all the KK states. 
The contribution from those states $m_N \ltap m_F$ is obviously finite. 
In the opposite limit $m_N \gg m_F$, 
$\Delta a^{(n)} _{\mu} \rightarrow (\alpha_X / 3 \pi)  m^2_F /m^2_N$. 
In this case the correction decouples as $m^{-2}_N$. Since the 
degeneracy of the KK states increases as $(m^2_N)^{p/2-1}$, however, 
the resultant sum is power divergent as $(m^2_N)^{p/2-1}$ for $p>2$, 
logarithmically divergent for $p=2$, and finite for $p=1$. 
As in the case of the bulk graviton loop, a 
 cutoff $\Lambda= M_*$ is used to regulate these divergences, and 
an overall factor of $\lambda \sim O(1)$ is inserted to parameterise 
the result of an actual one-loop computation in a more complete theory 
of quantum gravity. 
The contribution of those modes $m_N \gg m_F$ to $a_{\mu}$ is then, 
using a trivial modification of Eqn. \ref{cutoffsum},
\begin{eqnarray}
\label{gauge2}
\Delta a_{\mu} 
&=&\lambda \frac{\alpha_X}{3 \pi} m^2_F R^2 \left(\frac{\pi^2}{6}\right)
 \hbox{ }, \hbox{ }p=1 \\
\label{gauge3}
\Delta a_{\mu}
&=&2 \lambda \frac{\alpha_X}{3 \pi} 
\left(\frac{2}{p-2} \frac{\pi^{p/2} }{\Gamma[p/2]}\right)
\left(\frac{\bar{M}_{PL} m_F}{M^2_*}\right)^2
\left(\frac{M^p_* V_p}{M^n_* V_n} \right) 
\hbox{ }, \hbox{ } p>2 \\
\label{gauge5}
&=&2 \lambda \frac{\alpha^{(p)}_X}{3 \pi} 
\left(\frac{2}{p-2} \frac{\pi^{p/2} }{\Gamma[p/2]}\right)
\left(\frac{ m_F}{M_*}\right)^2  \hbox{ }, \hbox{ } p>2 \\
\label{gauge4}
\Delta a_{\mu}
&=&\lambda \frac{\alpha_X}{3 }  m^2_F R^2
  \ln  M^2_* /m^2_F \hbox{ }, \hbox{ }p=2.
\end{eqnarray}
To arrive at Eqn. \ref{gauge2}, $p=1$, I have neglected the mass 
of the zero mode. Also note that in this case the result is finite. 
For $p>1$ the symmetry breaking contribution is irrelevant since the
resulting sum is divergent. To get from Eqn. \ref{gauge3} to 
Eqn. \ref{gauge5}, I have used the 
relation $2\bar{M}^2_{PL}=M^{n+2}_* V_n$ and 
the relation Eqn. \ref{finestructure}
between the $4-$dimensional 
and $(4+p)-$dimensional fine structure constants. 

These expressions for the bulk vector boson contribution to $a_{\mu}$ 
can also be understood by considering computing the one--loop diagram 
in the full theory. 
Recall that the correction to the MMM requires a helicity flip in the loop 
diagram. This pulls out a 
factor of $p_{\alpha} p^a \gamma_a$, lowering the degree of divergence 
by two. Then  
naive power counting gives for $p>2$  
\begin{equation}
\Delta a_{\mu} \sim m^2_F \frac{\alpha^{(p)}_X}{M^p_*} \int dk k^{p+3} k^{-2} 
k^{-2} k^{-2} \sim m^2_F \frac{\alpha^{(p)}_X}{M^p_*} \Lambda^{p-2} .
\label{approx}
\end{equation}
The first two factors of $k^{-2}$ in the integrand 
are for the KK vector boson and 
fermion propagators.
Substituting $\Lambda \sim M_*$ further simplifies this result to 
$\Delta a_{\mu} \sim \alpha^{(p)}_X m^2_F /M^2 _*$, 
which is Eqn. \ref{gauge5}. 
Note that this result is independent of the dimension of the subspace.
This implies that the correction to MMM is independent of 
$V_p$, and depends only on 
$M_*$ and $\alpha^{(p)}_X$, the fine structure constant in the 
full theory. The dependence on $V_p$ only enters once 
the correction to the MMM 
is expressed in terms of the $4-$dimensional fine structure constant.  

{\it b. \underline{Applications}}

Requiring that $\Delta a_{\mu} <2 \times 0.8 \times 10^{-8}$  
gives the following 
$2 \sigma$ limits for the interesting cases $p=n=2$, and $p=n=6$:
\begin{equation}
p=n=2: \hbox{ } 
\alpha_X < 
3.3  \times 10^{-32} \times \frac{1}{\lambda}
\left(\frac{M_*}{\hbox{TeV}}\right)^4 
\hbox{ }, \hbox{ }p=n=6: \hbox{ } \alpha_X < 2.4 \times 10^{-31} 
\times \frac{1}{\lambda}\left(\frac{M_*}{\hbox{TeV}}\right)^4.
\label{glimit}
\end{equation}
The limit in the case $n=2$ is $\sim 10$ below the ``natural'' 
value of $\sim 2\times 10^{-31}$ for $M_*=$1 TeV. 
I emphasize that these 
tiny limits apply only to the situation where the 
vector boson occupies the whole bulk. The limits on $\alpha_X$
are considerably weaker if the vector boson lives in a 
subspace. That is, for other choices of $p< n$, 
the limit on $\alpha_X$ is  
weakened by the amount  
$M^n_* V_n /M^p_* V_p$. 
  
The limits in Eqn. \ref{glimit} translate into the following $2\sigma$
upper 
bounds on the mass of the zero mode, using $m_X < g_X M_*$:
\begin{eqnarray}
p=n=2: &&\hbox{ } m_X < 
1.7 \times 10^{-2} \times \frac{1}{\sqrt{\lambda}}
\left(\frac{M_*}{\hbox{3 TeV}}\right)^3 
\hbox{ } \hbox{eV} \hbox{ }, \\ 
p=n=6: &&\hbox{ } m_X <
4.7 \times 10^{-2} \times \frac{1}{\sqrt{\lambda}} 
\left(\frac{M_*}{\hbox{3 TeV}}\right)^3 
 \hbox{ } \hbox{eV}.
\label{masslimit}
\end{eqnarray}

In the case that $X=B-L$, the gauge boson can mediate an 
isotope dependent force at sub-mm distance scales. Even though 
the gauge coupling is microscopic, the force can still be huge 
compared to the gravitational force at those distance scales.
The limits in Eqn. \ref{glimit} quantify an upper limit 
to this signal. 
From the upper bound on the 
gauge boson mass, Eqn. \ref{masslimit}, it is clear that there is 
still an allowed region where $m_X \ll \hbox{mm}^{-1}$. Then the 
ratio of the forces is, for $p=n=2$ to $p=n=6$, 
\begin{equation}
F_X:F_{\hbox{grav}} =\alpha_X :G_N m^2_{\hbox{nucl}} \ltap
10^{6-7} \left(\frac{M_*}{\hbox{TeV}}\right)^4.
\end{equation}
The conclusion is that even though the $(g-2)$ constraint is
strong in the case $p=n$, the $B-L$ gauge boson can still mediate 
forces $10^6-10^7$ times stronger than gravity at sub-mm distances. 

Next I apply these results to the 
scenario of \cite{dienes}, where KK excitations of the SM gauge 
particles and Higgs bosons (and their superpartners) result in 
a power law evolution of the gauge couplings. This allows for a 
unification scale as low as few decades above the TeV scale. In this 
picture the SM gauge bosons, for example, have KK excitations in one or two 
extra dimensions. These will all contribute to the MMM. 
In what follows however, I only concentrate on the  
contribution from the KK photons for the reason that 
their contribution (and hence also those from KK $Z$'s and $W$'s) 
is rather small (see Eqn. \ref{mmmr} below). 
It is then seen that the constraint from $a_{\mu}$ provides a rather weak 
upper limit to $R$, the size of the subspace that the gauge 
bosons occupy.
Using Eqns. \ref{gauge2}, \ref{gauge4}, $\alpha^{-1}_{em}=137$,
$M_* \sim \hbox{10 TeV}$,  
 and the $2 \sigma$ experimental error 
on the muon $(g-2)/2$ gives 
\begin{equation}
p=1: \hbox{ }R^{-1} > \hbox{30 GeV} \hbox{ }, \hbox{ } 
p=2: \hbox{ }R^{-1} > \hbox{190  GeV}.
\label{mmmr}
\end{equation}
The limit for $p=1$ is finite, whereas the case $p=2$ is logarithmically 
sensitive to the cutoff.
Since these two constraints satisfy 
$R^{-1} \gg m_{\mu}$, the use of Eqns. \ref{gauge2} and 
\ref{gauge4} is consistent. The rather weak 
bound for $p=1$ is due to the 
$\alpha_{EM}$ suppression, and the constant density of KK states. 

Finally, KK excitations of the $Z_0$ gauge boson result 
in a tree--level correction to the theoretical predictions 
for atomic parity violation 
parameters. The measurement of atomic parity violation 
therefore constrain 
the size of 
some extra ``small'' dimensions. These experiments  
measure a quantity $Q_W$ which is the coherent sum of the $Z$ boson 
vector couplings to the nucleus. The measured value for Cesium is \cite{pdg}
\begin{equation}
Q^{exp}_W=-72.41 \pm 0.25 \pm 0.80 ,
\end{equation}
which agrees very well with the theoretical prediction \cite{pdg}
\begin{equation}
Q^{SM}_W=-73.12 \pm 0.06 \pm 0.01. 
\end{equation}
The exchange of a tower of KK $Z_0$ bosons trivially modifies (at tree-level) 
the SM 
prediction to\footnote{I assume that the SM fermions 
do not have any KK excitations below $M_*$.} 
\begin{equation}
Q^{SM}_W\rightarrow 
Q^{th}_W=Q^{SM}_W+Q^{KK}_W=Q^{SM}_W 
\left(1+ m^2_Z R^2 \sum_n \frac{1}{n^2}\right).
\end{equation}
The $m_Z$ contribution to the KK mass $n/R$ has been neglected.
Note that the KK correction to $Q_W$ 
pushes the theoretical prediction farther away from the measured value. 
To limit $R$, I require that $Q^{exp}_W-Q^{th}_W< 2 \sigma$, or  
$-Q^{KK}_W <0.89$. Using the measured value given above, and performing 
the sum over KK states results in the following $2 \sigma$ limits 
\begin{equation}
p=1:\hbox{ } R^{-1} > 1100 \hbox{ GeV} \hbox{ }; 
\hbox{ } p=2:\hbox{ }R^{-1}> 3.6 \hbox{ TeV} \sqrt{\frac{\ln M_*R}{\ln 20}}.
\label{atp}
\end{equation}
So it is quite clear that the stronger limits on TeV$-$sized extra dimensions 
are gotten from atmoic parity violation rather than the MMM. Compare 
the results in Eqn. \ref{atp} to Eqn. \ref{mmmr}.



\section{Bulk Right-Handed Neutrinos}

The authors of Ref. \cite{nima4} proposed several mechanisms for 
generating $\sim O(\hbox{eV})$ sized neutrino masses. 
In one of 
their models which is the attention of this section, small Dirac 
neutrino masses are generated by introducing bulk RH 
neutrinos. I now briefly explain their idea. The interaction 
\begin{equation}
\int d^4 x \hbox{ } k L H 
\sum_n v^{(n)}_c  
\label{rhn1}
\end{equation}
gives a Dirac mass $k v/\sqrt{2}$ to the LH and RH zero mode neutrino. 
The higher KK states also mix with the LH neutrino, but 
what will eventually concern us for the MMM are the extremely heavy KK 
states, which have very tiny mass mixing with the LH neutrinos. 
In their model 
this interaction arises from a higher dimension operator in the 
full theory. More specifically, the operator 
\begin{equation}
\int d^4 x d^p y \frac{k_0}{M^{p/2} _*} L(x,y)
H(x,y) \frac{v^c(x,y)}{\sqrt{V_p}},    
\end{equation}
where I have allowed the fields to propagate in $p$ extra dimensions, 
and $k_0$ is a dimensionless constant. Inserting the approximation 
$\psi _{SM} (x,y) \sim \sqrt{\delta^{(p)}(y)} \psi _{SM} (x)$ for 
the SM zero modes \cite{nima4}, and the usual 
KK mode expansion for $v^c$ gives the relation between $k$ and 
$k_0$:
\begin{equation}
k=k_0  \frac{1}{\sqrt{M^p_* V_p}}=
k_0 \left(\frac{M_*}{\bar{M}_{PL}}\right)^{p/n},
\end{equation}
where to obtain 
the second relation I used $(4 \pi G_N)^{-1}=M^{n+2}_* R^n$ and 
assumed $V_p=R^p$ for simplicity. Consequently, the coupling $k$ can be 
very tiny, and neutrino masses of the correct order of magnitude 
required 
to explain the atmospheric and solar neutrino anomalies can  
be obtained. 

To obtain a correction to the MMM at the one-loop level, I assume:

i) the muon neutrino obtains a Dirac mass by mixing with a sterile 
neutrino, as in Eqn. \ref{rhn1}; 

ii) the Higgs sector is extended 
to include an extra Higgs doublet which also acquires a vev. 

The second assumption is realised in low-energy supersymmetric 
extensions to the SM, for example. The point of this assumption 
is to obtain a trilinear 
coupling between the physical charged 
Higgs, the muon, and the bulk RH neutrinos. Consequently, these 
interactions contribute to the MMM at one-loop, as 
shown in Fig. \ref{extradfig2}b. 

An evaluation of Fig. \ref{extradfig2}b gives the following correction 
to the MMM:
\begin{equation}
\Delta a^{(n)}_{\mu}=\frac{2 k^2}{16 \pi^2} \int dx \frac{m^2_F}
{\tilde{L}(x,m^2_F,m^2_{\phi^+},m^2_N)} x^2 (1-x),
\end{equation}
where I have denoted by $k$ the trilinear coupling of the physical 
charged Higgs, charged LH fermion, and bulk RH neutrinos. It is related 
to a linear combination of 
the trilinear couplings in the gauge basis. 
The function $\tilde{L}$ is 
$\tilde{L}(x,m^2_F,m^2_{\phi^+},m^2_N)
=-x(1-x) m^2_F+x m^2_{\phi^+}+(1-x) m^2_N$, where $m_{\phi^+}$ is 
the mass of the charged Higgs. In the limit 
$m^2_N \gg m^2_F, m^2_{\phi^+}$,
\begin{equation}
\Delta a^{(n)}_{\mu}=\frac{2}{3} \frac{ k^2}{16 \pi^2} \frac{m^2_F}{m^2_N}. 
\label{rhn3}
\end{equation}   

As in the case of the bulk gravitons and bulk vector bosons, the 
dominant correction to the MMM occurs from the superheavy KK RH neutrinos. 
In fact, the sum over KK states is identical (up to numerical factors) 
to the KK sum for the vector bosons. 
Integrating Eqn. \ref{rhn3} for $p>2$, and using trivial 
modifications to Eqns \ref{fR} and \ref{cutoffsum}, gives 
\begin{equation}
\Delta a_{\mu}=\frac{2}{3}k^2  \frac{ \lambda}{8 \pi^2} 
\frac{2}{p-2} \frac{\pi^{p/2}}{\Gamma[p/2]} \frac{m^2_F}{M^2_*} 
\left(\frac{\bar{M}_{PL}}{M_*}\right)^{2p/n}. 
\end{equation}
Note that since $k=k_0 (M_*/M_{PL})^{p/n}$, the strong dependence 
on $M_*/\bar{M}_{PL}$ cancels, leaving
\begin{equation}
\Delta a_{\mu}=\frac{2}{3}k_0 ^2  \frac{ \lambda}{8 \pi^2}
\frac{2}{p-2} \frac{\pi^{p/2}}{\Gamma[p/2]} \frac{m^2_F}{M^2_*}.
\end{equation}
Choosing 
$p=5$, for example, gives 
\begin{equation}
\Delta a_{\mu} \sim 8 k^2_0 
\lambda \times 10^{-10} \left(\frac{\hbox{TeV}}{M_*}\right)^2 
\sim k^2_0 \lambda \times 10^{-9}  \left(\frac{\hbox{TeV}}{M_*}\right)^2. 
\end{equation}
So since both $k_0$ and $\lambda \sim O(1)$,
 it is possible to get $\Delta a_{\mu}\sim 10^{-9}$. This 
may be within the reach of the BNL E821 experiment. 

\section{Acknowledgements}

The author would like to thank N. Arkani-Hamed and M. Suzuki 
for discussions and 
encouraging me to complete this project. The author also 
thanks T. Moroi for useful communications about the BNL E821 
experiment.  
This work was supported in part by the Director, Office of Energy Research, 
Office of High Energy and Nuclear Physics, Division of High Energy Physics 
of the U. S. Department of Energy under Contract DE-AC03-76SF00098.
The author also thanks the Natural Sciences and Engineering Research 
Council of Canada (NSERC) for their support. 

\end{document}